# Direct Observation of nm-Scale Mg- and B-Oxide Phases at Grain Boundaries in $MgB_2$

R. F. Klie, J. C. Idrobo and N. D. Browning

*Department of Physics (M/C 273), University of Illinois at Chicago, 845 West Taylor Street, Chicago, IL 60607-7059*

K.A. Regan, N.S. Rogado, and R. J. Cava

*Department of Chemistry and Princeton Materials Institute, Princeton University, Princeton, N J, 08544*

Here we describe the results of an atomic resolution study of the structure and composition of both the interior of the grains, and the grain boundaries in polycrystalline $MgB_2$. We find that there is no oxygen within the bulk of the grains but significant oxygen enrichment at the grain boundaries. The majority of grain boundaries contain $BO_x$ phases smaller than the coherence length, while others contain larger areas of MgO sandwiched between $BO_x$ layers. Such results naturally explain the differences in connectivity between the grains observed by other techniques.

The discovery of superconductivity in $MgB_2$, with a record breaking transition-temperature $(T_c=39K)$[1] for "conventional" superconductors, has sparked an enormous research effort to understand its structure-property relationships. One aspect of $MgB_2$ that has particular technological importance is that the grain boundaries in the polycrystalline materials studied so far do not appear to significantly decrease the overall critical current $(J_c)$.[2,3] This is in marked contrast to the cuprate high-$T_c$ superconductors, where high-angle grain boundaries universally act as weak links and dramatically reduce the inter-granular critical currents[4,5], and offers the promise of a vast number of new industrial applications. A critical first step in understanding the mechanism by which these boundaries can support a high critical current is a determination of the structure and composition of both the grains and the grain boundaries. Here we report a series of direct atomic resolution[6] studies of the grains and the grain boundaries in a polycrystalline $MgB_2$ sample. Specifically we find that there is no oxygen within the bulk of the grains (within the few percent detection limits of our techniques) but significant oxygen enrichment at the grain boundaries. Furthermore, the boundaries are found to consist of two distinct boundary types: one containing $BO_x$ phases with a width of <4nm (i.e. smaller than the coherence length) and a second containing a $BO_x$ -$MgO_y(B)$ - $BO_z$ trilayer ~10-15nm in width (i.e. larger than the coherence length). Such boundary features indicate that although $J_c$ is high overall, the structure-property relationships at grain boundaries in $MgB_2$ are a complex function of processing conditions, and that control of the oxygen content at grain boundaries will be essential for attaining optimal bulk critical currents.

The samples used in this study were synthesized from powders of Mg and B, which were sintered and hot-pressed into pellets (for a more detailed description of the specimen preparation see ref. [2]). The samples were then tested for their superconducting behavior and a transition temperature of 37-38 K was determined.[2] Previous bulk samples prepared in the same manner were investigated by magneto-optical techniques[3], which highlighted extensive inhomogeneities in the magnetic flux penetration across the sintered pellet, i.e. the grain boundaries showed much higher magnetic flux penetration at low fields with some being much better connected than others. Inhomogeneities in the structure and composition of $MgB_2$ pellets have also been seen by X-ray diffraction (XRD) and low-resolution energy dispersive X-ray spectroscopy (EDS). In particular, secondary phases (~ 5%) of MgO [3] were observed and the presence of $MgB_4$ impurities at the interfaces was inferred from a decrease in the characteristic Mg X-ray signal.[7]

Correlating the observed flux penetration at the interfaces with the observed structural or compositional changes is made difficult by the fact that the grain size of $MgB_2$ is on the order of tens of nanometers. As such, there are few analytical techniques that can directly isolate measurements on the grain boundaries from those on the grains. One means to directly observe structure, composition and bonding effects at grain boundaries on the atomic scale is through the combination of Z-contrast imaging and electron energy loss spectroscopy (EELS) at atomic resolution in a scanning transmission electron microscope (STEM)[6]. Our experiments here use a 200kV JEOL 2010F STEM which has a point resolution of ~0.14nm in scanning mode[8]. In this experimental condition, the incoherent high-angle annular dark field or "Z-contrast" image[9] allows the structure of the grains/grain boundaries to be directly observed and the image can also be used to position the electron probe for (EELS)[6]. Core loss EELS probes the unoccupied density of states near the conduction band minimum and is exactly analogous to near edge X-ray absorption spectroscopy (XAS), but with a much higher spatial resolution afforded by an electron microscope. In the experiment performed here, the EEL spectra are acquired directly from the grain boundaries with characteristic core losses used to determine composition and spectral fine-structure used to identify order in the structure (crystallinity). The samples were prepared for microscopy by the standard TEM specimen techniques of mechanical grinding (using the wedge technique) followed by ion-milling with 3 kV argon ions until electron transparency. The samples were Plasma cleaned with Ar-ions prior to insertion in the microscope.

Initial evaluation of the microstructure by low magnification Z-contrast images confirms the expected small grain size in this material; with grains ranging in size from 20-400 nm. Figure 1 (a) shows a high magnification image of the atomic structure of one of the grains in the [010] orientation. The bright dots in the unfiltered Z-contrast image correspond to the Mg columns. The low signal to noise ratio of this imaging technique and the low scattering amplitude of boron (the image intensity is approximately proportional to $Z^2$) means that the central two B atomic columns are not directly visible in the raw image. However, the structure is confirmed to be bulk $MgB_2$ from the convergent beam electron diffraction (CBED) pattern and the Fourier filtered image (see inset) begins to resolve the B columns (the intensities inside the Mg rectangle can also be formed by the tails of the electron probes; further image simulations are necessary). Confirmation that these columns are representative of the crystal structure comes from the intensity ratio, which shows good agreement to the expected scattering amplitudes of Mg and B columns. It is interesting to note that in the grains studied we could not detect atomic scale bulk defects such as stacking faults or twin boundaries that were reported by other groups[7]. Furthermore, the electron energy loss spectra in Figure 1(b) show that the central regions of the grains do not contain significant levels of oxygen. As the sensitivity of EELS is ~3-5%[10], we cannot definitively rule out the presence of oxygen in the centers of the grains, but the images spectra and CBED patterns strongly suggest that the oxygen content is very low.

Figure 2 shows one of the larger grains we observed and several interesting structural features associated with the grain boundaries. The grain boundaries appear as dark lines between the individual grains, which indicates a secondary phase or material with a different composition. Note that strain has been known to cause contrast changes at grain boundaries [11,12], but in this case the contrast did not change with collection conditions and EELS confirms the presence of a secondary phase. The feature labeled A in figure 2 (a) shows two dark lines, which surround an unusually shaped platelet. This structure is shown at higher magnification in figure 2 (b). These platelets can easily be identified from lower magnification images and appear to be present at a significant proportion of the grain boundaries. EEL spectra, acquired from a position inside the bulk of grains, from the grains close to the dark line interface, the dark line at the interface and the intergrowth layer (feature A in Figure 2(a)) are shown in figure 2 (c). The material close to the internal interface does not show any difference in the image from the bulk structure, and similarly the EELS from this region and the bulk are the same.

The dark interfacial layer on the other hand shows enormous changes in the EELS spectra compared to the bulk features. The main features of the spectrum show that this dark interfacial region is composed mainly of boron and oxygen in some compound form $BO_x$ which is typically ~2 nm thick. The lack of crystalline contrast suggests that the material is amorphous. The EEL spectra also show that the platelets between the two dark layers of $BO_x$ are of a completely different composition. Here the B K-edge intensity (figure 2(c)) is reduced dramatically and the oxygen K-edge is increased by a similar amount. The fine structure of the oxygen peak and the sensitivity of the platelet to beam damage [13] is consistent with crystalline MgO. The thickness of these MgO phases is typically ~10nm in the boundaries observed. This particular type of grain boundary appears to therefore be a tri-layer containing phase separated $BO_x$ -$MgO_y(B)$ - $BO_z$. The presence of B in the MgO platelet cannot be exclusively ruled out for the same sensitivity reasons described above.

At the grain boundaries where these MgO platelets are absent, a different situation is found. Similar to the type of boundary described above, a $BO_x$ layer is observed at the boundaries. However, as can be seen from the Z-contrast image in figure 3, there is a structural change in the grains adjacent to the $BO_x$ layer. The energy loss spectra are not consistent with bulk MgO material in this region, but analysis of the edge intensity (taking into account the scattering cross-section) indicates that the O content in this layer is increased to 11%. Again these results indicate a significant uptake of oxygen at the grain boundaries. For grain boundaries that show these features, the width of the oxygen enriched layer is typically ~1-2 nm.

The results described above clearly show that the grain boundaries in $MgB_2$ exhibit a propensity for oxygen enrichment. At this stage it is not clear whether the two different types of boundaries we observe are fundamentally different or simply different stages of grain boundary oxidation. Conceivably the initial stages of oxidation, i.e. low oxygen content, could result in the formation of the $BO_x$ layers at the boundary, bordered by the oxygen rich $MgB_2$ layers. As oxidation continues, i.e. the oxygen content in the sintered pellet increases, the oxygen content of the $MgB_2$ layers increase until phase separation occurs and MgO nucleates. Once nucleated, the MgO grain boundary precipitates could continue to grow with increasing oxygen content and thereby result in the second phases observed in X-ray diffraction. Such results would naturally explain the differences in connectivity between the grains observed by magneto-optical imaging[2]. It is not immediately clear exactly how each type of grain boundary layer affects transport. However, what is clear from this analysis is that the width of the majority of grain boundaries (i.e. the type 2 boundaries) is ~1 – 3 nm, which is smaller than the superconducting coherence length[2] and by itself may be enough to explain the high inter-granular critical currents in $MgB_2$[3].

This work is supported by the US Department of Energy under grant number DOE FG02 96ER45610. The experimental results were obtained on the JEOL 2010F operated by the Research Resources Center at UIC and funded by NSF.

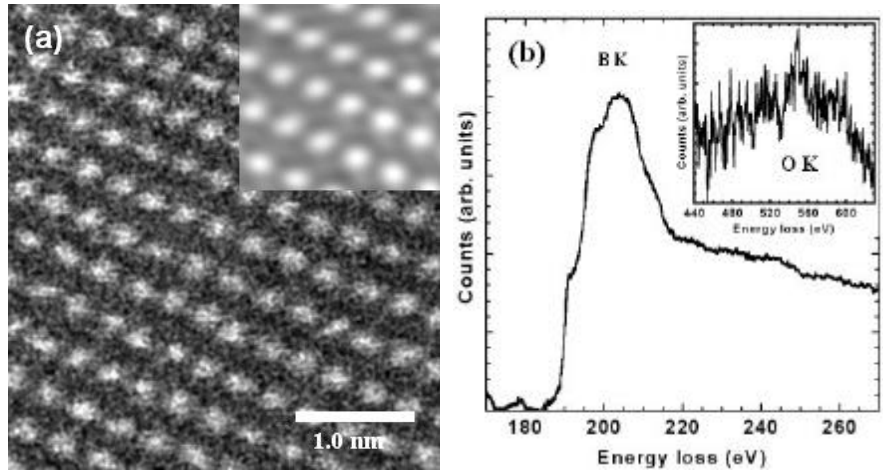

FIG 1. a) High resolution Z-contrast image of bulk $MgB_2$ in the [010] orientation. The bright spots represent the Mg atoms in the rectangular unit cell. The B atoms are not clearly visible in this image, but the inset shows a Fourier-filtered section of this image and here the intensities inside the Mg-rectangles could indicate the boron positions.
b) EEL spectra from the bulk of the $MgB_2$ grains showing the boron K-edge with a dispersion of 0.2 eV/channel and an acquisition time of 3 s per spectrum. A pre-peak, as well as a splitting of the main peak can be estimated. The inset shows a magnified spectrum from the energy loss of oxygen. Clearly no oxygen signal is visible here.

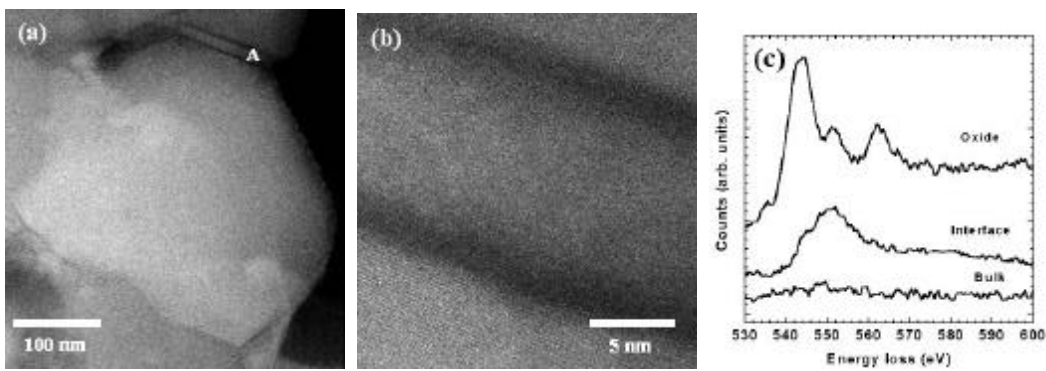

FIG 2. a) Z-contrast image of a single-phase grain. The feature labeled A indicates the presence of an MgO layer. The two dark lines on top and below this feature represent the boron and oxygen rich phases as confirmed by EELS.

b) Enlarged micrograph of the boundary area in figure 1. This image shows an even distribution of image contrast in the bulk of the grain, indicating no segregation of oxide towards the interface. The low contrast in the interface layer suggests an Mg depleted phase, consisting possibly of $BO_x$.

c) EEL spectrum taken from the bulk, the dark feature as well as the oxide layer with a dispersion of 0.5 eV per channel. The acquisition time for each spectrum was 3 s and each spectrum is a background subtracted, deconvoluted sum of 6 individual spectra. The spectra are magnified by a factor of 8 to show the O K-edge fine structure.

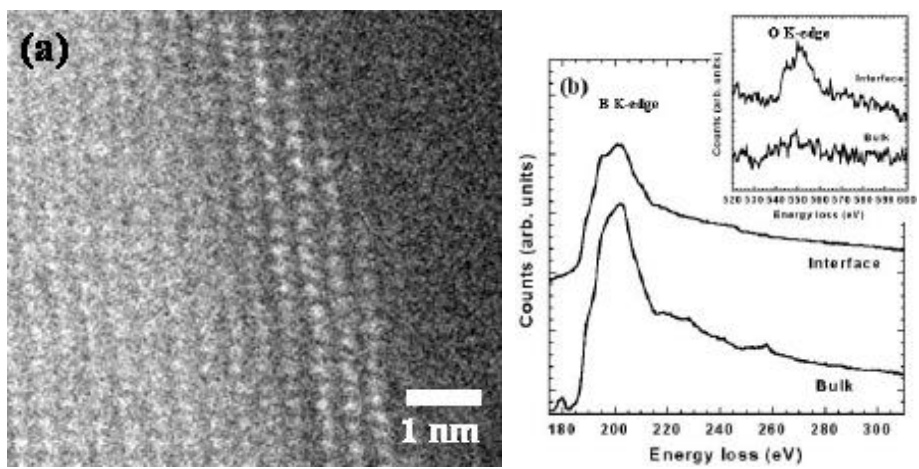

FIG 3. a) High resolution Z-contrast image of a different grain boundary. This image shows the familiar bulk and dark interface features. A 3-4 unit cell wide layer of brighter contrast, which was identified by EELS to be an $MgB_xO_y$ layer can be clearly observed.

b) EEL spectra from the bulk and the bright interface layers showing the boron K-edges at an energy onset of 188 eV with a dispersion of .05 eV/channel and an acquisition time of 3s. The inset reflects the changes in the O K-edge structure from the bulk to the interface. The oxygen spectra are both magnified by a factor of 8 to show the fine-structure.